\documentclass[prl,aps,showpacs,floatfix,superscriptaddress,twocolumn,nofootinbib,citeautoscript]{revtex4-1}

\usepackage{amsmath}
\usepackage{amsfonts}
\usepackage{amssymb}
\usepackage{revsymb}
\usepackage{graphicx}
\usepackage{siunitx}
\usepackage{bm}
\usepackage{dcolumn}
\usepackage{multirow}
\usepackage{graphicx}
\graphicspath{{figures/}}
\usepackage{epsfig}
\usepackage{float}
\usepackage{tabularx}
\usepackage{booktabs}
\usepackage{xcolor}

\usepackage{pdfpages}
\usepackage{etoolbox} 
\makeatletter
\patchcmd{\@outputpage@head}{\@ifx{\LS@rot\@undefined}{}{\LS@rot}}{}{}{}
\makeatother

\def\ket#1{|\,#1 \,\rangle}

\def\mT{\,\mathrm{mT}}

\def\Hz{\,\mathrm{Hz}}

\def\fref#1{Fig.~\ref{#1}}

\begin{document}
\title{$^{176}$Lu$^+$ clock comparison at the $10^{-18}$ level via correlation spectroscopy}
\author{Zhang Zhiqiang}
\affiliation{Centre for Quantum Technologies, 3 Science Drive 2, 117543 Singapore}
\author{K. J. Arnold}
\affiliation{Centre for Quantum Technologies, 3 Science Drive 2, 117543 Singapore}
\affiliation{Temasek Laboratories, National University of Singapore, 5A Engineering Drive 1, 117411 Singapore}
\author{R. Kaewuam}
\affiliation{Centre for Quantum Technologies, 3 Science Drive 2, 117543 Singapore}
\author{M. D. Barrett}
\email{phybmd@nus.edu.sg}
\affiliation{Centre for Quantum Technologies, 3 Science Drive 2, 117543 Singapore}
\affiliation{ Department of Physics, National University of Singapore,  2 Science Drive 3, 117551 Singapore}
\affiliation{National Metrology Centre, Agency for Science, Technology and Research (A*STAR), 8 CleanTech Loop, Singapore 637145}
\begin{abstract}
We experimentally demonstrate agreement between two $^{176}$Lu$^+$ frequency references using correlation spectroscopy.  From a comparison at different magnetic fields, we obtain a quadratic Zeeman coefficient of  $-4.89264(88)\,\Hz/\mT^2$, which gives a corresponding fractional frequency uncertainty contribution of just $2.5\times 10^{-20}$ for comparisons at typical operating fields of 0.1\,mT.   A subsequent comparison with both systems at 0.1\,mT, demonstrates a fractional frequency difference of $(-2.0\pm(3.7)_\mathrm{stat}\pm(0.9)_\mathrm{sys})\times10^{-18}$, where `stat' and `sys' indicate statistical and systematic uncertainty, respectively.
\end{abstract}
\maketitle
The last two decades has seen rapid development of optical atomic clocks, with many systems now reporting fractional performance at the $10^{-18}$ level~\cite{SrYe2,huang2021liquid,mcgrew2018atomic,YbPeik,AlIon,takamoto2020test,ushijima2015cryogenic}.  Such performance has many important applications, such as defining the geoid~\cite{lion2017determination}, monitoring geopotential changes~\cite{bondarescu2015ground}, measuring time and frequency~\cite{leroux2017line}, and testing fundamental physics~\cite{safronova2018search}.   Independent of the proposed application, it is important to explore a wide range of atomic systems to assess their merits and long-term potential for future development.  In this work we demonstrate the ease at which two $^{176}$Lu$^+$ references can achieve agreement at $\lesssim10^{-18}$.  From a comparison at different magnetic fields, we obtain a quadratic Zeeman coefficient of  $-4.89264(88)\,\Hz/\mT^2$, which gives a corresponding fractional frequency uncertainty contribution of just $2.5\times 10^{-20}$ for comparisons at typical operating fields of 0.1\,mT.   A subsequent comparison with both systems at 0.1\,mT, demonstrates a fractional frequency difference of $(-2.0\pm(3.7)_\mathrm{stat}\pm(0.9)_\mathrm{sys})\times10^{-18}$, where `stat' and `sys' indicate statistical and systematic uncertainty, respectively.  Due to the clock transition's insensitivity to external electromagnetic perturbations, results were obtained without magnetic shielding \cite{dube2013evaluation} or active field stabilization, without ground state cooling using simulations to justify its effectiveness \cite{chen2017sympathetic}, without extreme temperature measurement and control \cite{SrYe2,mcgrew2018atomic,middelmann2012high,huang2021liquid}, or even the use of simulations and thermal cameras for a more moderate temperature assessment \cite{dolevzal2015analysis,YbPeik}.  Such ease of use offers considerable technical advantages for practical implementations of an atomic frequency reference.

\begin{figure}
\includegraphics[width=0.49\textwidth]{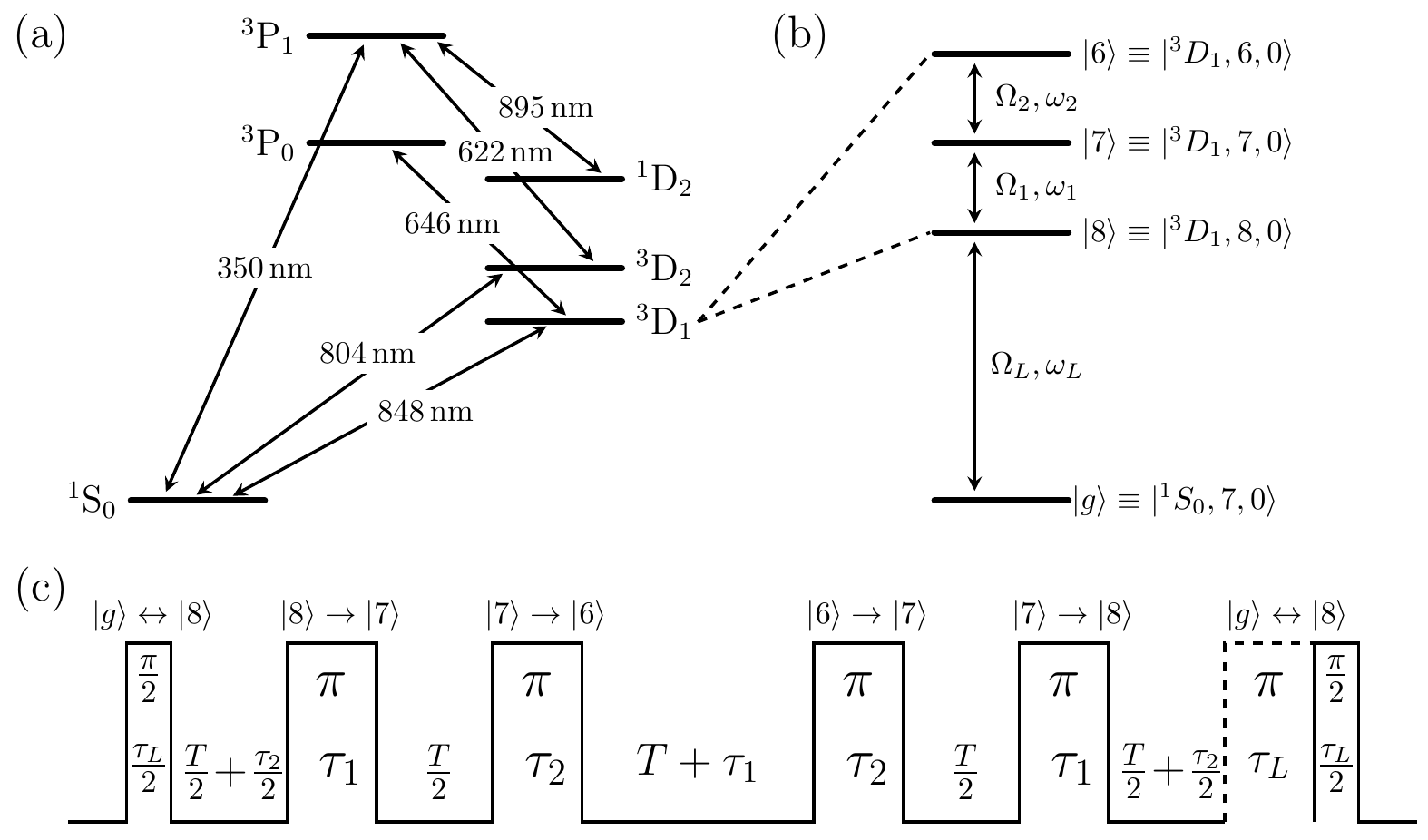}
\caption{$^{176}$Lu$^+$ spectroscopy (a) Atomic level structure of $^{176}$Lu$^+$ showing the wavelengths of transitions used. (b) Levels of the 848-nm clock transition used in the clock interrogation sequence.  $\Omega_\alpha$ and $\omega_\alpha$ denote the coupling strengths and frequencies for the fields driving the transitions indicated. (c) Clock interrogation sequence for HARS.  An optional $\pi$-pulse on the optical transition is included when implementing hyper-Ramsey spectroscopy to suppress probe-induced ac-Stark shifts.}
\label{fig:scheme}
\end{figure}

Singly ionized lutetium is a two electron system with three available clock transitions and the relevant level structure is shown in \fref{fig:scheme}a.  As described in previous work~\cite{zhiqiang2020hyperfine,kaewuam2020precision}, laser beams at 350, 622, and 895\,nm facilitate optical pumping into the long-lived $^3D_1$ state, laser beams at 646\,nm facilitate detection, Doppler cooling, and state preparation of $\ket{7}\equiv\ket{^3D_1,7,0}$, and microwaves are used to drive transitions between the $^3D_1$ hyperfine states.  This work concerns the $^1S_0\leftrightarrow{}^3D_1$ transition at 848\,nm, which has favourable properties as a frequency reference.  The blackbody radiation (BBR) shift is $\approx -1.36(10)\times 10^{-18}$ at 300\,K \cite{arnold2018blackbody,arnold2019dynamic}, which is sufficiently small that any temperature-induced frequency difference between two references operating within a temperature range of $25\sim60\,{}^\circ\mathrm{C}$ is $\lesssim 10^{-18}$.  At a typical operating magnetic field of $0.1\,\mathrm{mT}$, transitions between the $m=0$ states have the lowest sensitivity to magnetic field noise of any existing clock system, being at least five-fold smaller than those in neutral lattice clocks \cite{SrYe2,mcgrew2018atomic}, a 100-fold smaller than in Al$^+$ \cite{rosenband2007observation}, and five orders of magnitude smaller than those in the alkaline-earth ions (Ca$^+$ and Sr$^+$)\cite{dube2013evaluation,cao2017}.  The large atomic mass makes it less susceptible to motional shifts than lighter ions and the secondary clock transition at 804\,nm provides exceptional detectability of micromotion.  Owing to the upper-state lifetime of approximately one week, the dominant systematic is the probe-induced ac Stark shift, which is 200-fold smaller than for the Yb$^+$ E3 clock transition under the same interrogation conditions \cite{arnold2018blackbody,arnold2019dynamic,hyperRamsey2012peik}.

All clock systems use some form of averaging, which is typically an averaging over one or more pairs of Zeeman lines \cite{bernard1998laser} or over orientations of the magnetic field \cite{itano2000external}.  In the case of $^{176}$Lu$^+$, we use an averaging over transitions from $\ket{g}\equiv\ket{^1S_0,7,0}$ to the hyperfine states $\ket{F}\equiv\ket{^3D_1,F,0}$ where $F=6,7,$ and $8$ \cite{MDB1}.  The average can be conveniently realized using a hybrid microwave and optical interrogation~\cite{kaewuam2020hyperfine}, which we refer to here as hyperfine-averaged Ramsey spectroscopy (HARS).  As illustrated in \fref{fig:scheme}b\&c, the interrogation sequence consists of a Ramsey interrogation on the $\ket{g}\leftrightarrow\ket{8}$ transition and microwave transitions within the interrogation time to transfer population between the hyperfine states.  Timing is such that the effective time in each of the hyperfine states is the same.  When the laser is servoed to the central Ramsey fringe, the hyperfine-averaged (HA) clock frequency is given by 
\begin{equation}
f = \frac{1}{3}(f_{8}+f_{7}+f_{6})=f_L+\frac{1}{3}(2f_1+f_2),
\label{Eq:clockfreq}
\end{equation}
where $f_{F}$ is the resonant frequency of the $\ket{g}\leftrightarrow\ket{F}$ transition, $f_L=\omega_L/(2\pi)$ is the laser frequency, and $f_k=\omega_k/(2\pi)$ for $k=1,2$ are the microwave frequencies used within the interrogation sequence.  As explained in previous work~\cite{kaewuam2020hyperfine}, the interrogation sequence ensures $f_L+(2f_1+f_2)/3$ represents the HA clock frequency, as indicated by Eq.~\ref{Eq:clockfreq}, even if the microwave fields are not exactly resonant.

Comparisons were carried out on two ions held in separate traps, denoted Lu-1 and Lu-2, the details of which are reported elsewhere \cite{arnold2020precision,kaewuam2020precision}.  As our laser coherence limits interrogation times to $\sim 50\,\mathrm{ms}$, we use correlation spectroscopy \cite{chwalla2007precision,chou2011quantum,clements2020lifetime}, which allows extended probe times by factoring out laser noise.  In our case we have been able to extend interrogation times to over 700\,ms, which is limited only by a heating problem in one trap (Lu-1).    Each comparison consisted of three interleaved experiments: one to measure the frequency difference, $\delta f$, between the two atomic references using HARS; one to keep the 848-nm clock laser near to resonance with the $\ket{g}\leftrightarrow \ket{8}$ transition using Rabi spectroscopy; and one to monitor the magnetic field in each chamber using microwave spectroscopy of the $\ket{{}^3D_1,7,0}\leftrightarrow \ket{{}^3D_1,6,\pm1}$ transition.  The duty cycle for measuring the frequency difference was $87\%$ for the longest interrogation times used in the experiments.

Typical systematic effects for the comparison experiments are given in table~\ref{tab:system}.  The dominant systematic shift arises from the probe-induced ac-Stark shift.  This was measured in each trap by interleaving Rabi and hyper-Ramsey spectroscopy \cite{yudin2010hyper,hyperRamsey2012peik} of the $\ket{g}\leftrightarrow\ket{8}$ transition with the frequency difference determining the probe-induced ac-Stark shift, $\delta_k$, for Lu-$k$.  For HARS, the resulting shift given in table~\ref{tab:system} is~\cite{yudin2010hyper}
\begin{equation}
\delta_{s,k} = \frac{\delta_{k}}{1+\frac{\pi}{2}\frac{T_R}{\tau_L}},
\label{Eq:848stark}
\end{equation}
where $T_R=3(T+\tau_1+\tau_2)$ is the total interrogation time and $\tau_L$ is the $\pi$-pulse time for the clock transition.  Several ac-Stark shift measurements were taken throughout the campaign and the $\delta_{k}$ for each experiment is determined from measurements of nearest proximity in time. As the shift is not continuously monitored, we assumed a 5\% uncertainty in the ac-Stark for each trap to allow for possible variation during the experiment but bound by the maximum variation observed between measurements. The value of $\delta_{k}$ given in table~\ref{tab:system} is a single representative example.
\begin{table*}
\caption{Characteristic uncertainty budget. All values are relative to $10^{-18}$ of the HA frequency. The values given here are for experiment 7 for which $\tau_L=8\,\mathrm{ms}, \tau_1=\tau_2=24\,\mathrm{ms},$ and $T=200\,\mathrm{ms}$ with an applied static magnetic field of $\approx 0.1\,\mathrm{mT}$ for both traps. Tables for experiments 1 to 6 are given in the supplemental material.}
\label{tab:system}
\begin{tabular*}{\textwidth}{@{\extracolsep{\fill}} >{\quad}l r r r r r r <{\quad}}
\toprule[0.75pt]
 & \multicolumn{2}{c}{Lu-1} &  \multicolumn{2}{c}{Lu-2} & \multicolumn{2}{c}{Difference} \\
\cmidrule[0.5pt](lr){2-3} \cmidrule[0.5pt](lr){4-5} \cmidrule[0.5pt](r{0.75em}){6-7}
Effect & \multicolumn{1}{c}{Shift} & \multicolumn{1}{c}{Unc.} & \multicolumn{1}{c}{Shift} & \multicolumn{1}{c}{Unc.} & \multicolumn{1}{c}{Shift} & \multicolumn{1}{r<{\quad}}{Unc.} \\  \hline
Excess micromotion &-0.41&0.37 &-0.44 &0.34&0.03 &0.50\\
Second-order Doppler (thermal)&-1.87&0.45&-0.13&0.06&-1.75&0.45\\
ac-Zeeman (rf)&0.54&0.01&0.15&0.01&0.39&0.01\\
ac-Zeeman (microwave)&-0.06&0.04&-0.06&0.15&-0.01&0.16\\
Gravity shift\footnote{Only differential shifts are considered.}&-&-&-&-&-1.31&0.15 \\
ac-Stark shift&-126.59&6.33&-125.14&6.26&-1.45&8.90\\
HARS timing&2.60&1.33&4.88&0.72&-2.28&1.51\\
Microwave coupling&0&0.21&0&0.21&0&0.30\\
Optical coupling&0&0.28&0&0.28&0&0.39\\
Residual quadruple shift&0.22&0.02&0&0.32&0.22& 0.32\\
\textbf{Total shift} &-125.58&6.50&-120.74&6.33&-6.15&9.07\\
\bottomrule[0.75pt]
\end{tabular*}
\end{table*}

Excess micromotion (EMM) was evaluated on the $\ket{^1S_0,7,0}\leftrightarrow\ket{^3D_2,9,0}$ transition at 804\,nm by the resolved sideband technique \cite{berkelandMicro} and several measurements were carried out throughout the campaign.  For each trap, the mean and half the difference of the largest and smallest determinations are given for the shift and uncertainty, respectively.  Thermal motion was investigated from dephasing when driving the $\ket{g}\leftrightarrow \ket{8}$ transition and variable delays before probing were used to estimate heating rates.  The associated second-order Doppler shift is dominated by a high heating rate in Lu-1 and the fractionally larger uncertainty arises from the accuracy at which that could be assessed.  The heating rates were 2.3(0.8)$\,\mathrm{mK/s}$ and 0.099(15)$\,\mathrm{mK/s}$ for Lu-1 and Lu-2, respectively.  These were measured repeatedly throughout the campaign, with the larger heating rate being measured almost daily.

Time-varying magnetic fields give rise to an ac-Zeeman shift with the dominant contributions coming from currents in the electrodes driven by the rf-trapping potential.  The contribution depends on the component of the rf-magnetic field perpendicular to the applied dc field~\cite{gan2018oscillating}.  This was measured from an Autler-Townes splitting on the Ba$^+$ clock transition~\cite{arnold2020precision}, giving rms values of $0.98(1)\,\mathrm{\mu T}$ and $0.52(1)\,\mathrm{\mu T}$ for Lu-1 and Lu-2 respectively.  

The microwave fields used in HARS can also contribute to an ac-Zeeman shift~\cite{kaewuam2020hyperfine}.  Inasmuch as practically possible, microwave antennas are positioned to maximise the $\pi$ component and balance the $\sigma^\pm$ components, which suppresses the contribution to $\lesssim 7 \times 10^{-19}$ for any of the comparisons reported here.  The microwave field components are deduced from measured $\pi$-times on the $\ket{7,0} \leftrightarrow \ket{F,m_F=0,\pm1}$ where $F=6,8$ and the uncertainty given in the table is based on a 5\% inaccuracy for the measured $\pi$-times.  Detailed calculations are given in the supplemental material.

The hyperfine interaction gives rise to a quadrupole shift that is not eliminated by HA.  The shift is characterized by a residual quadrupole moment of $-2.5\times 10^{-4} e a_0^2$ as determined from a combination of $g$-factor measurements and theory \cite{zhiqiang2020hyperfine}.  The shift can then be estimated using the quadrupole shift extracted from $S_{1/2}\leftrightarrow{D_{5/2}}$ clock measurements in Ba$^+$ and rescaling the quadrupole moments.  This has been done for Lu-1 \cite{arnold2020precision}, which gives an estimated shift of $2.2\times 10^{-19}$ with a 10\% uncertainty ascribed to the value of the quadrupole moment.  For Lu-2, the maximum possible shift from the dc confinement would be $3.2\times 10^{-19}$ if the magnetic field was aligned along the trap axis.  For both traps, the magnetic field is at an angle of approximately $57^\circ$ to the trap axis.  As such, we would expect a reasonable suppression of the component from the trapping potentials and the shift dominated by stray fields.  As it is unlikely a shift from stray fields would exceed the maximum possible shift from the dc confinement, we bound the shift by $\pm 3.2 \times 10^{-19}$.

In the analysis of HARS \cite{kaewuam2020hyperfine} it was assumed the detuning of the microwaves was much smaller than the associated coupling strengths, and the finite duration of the optical pulses was neglected.  For the level of precision obtained here, these assumptions should be reconsidered.  A comprehensive analysis is given in the supplemental material.  Most importantly, the finite duration of the optical pulse is associated with an effective time $2\tau_L/\pi$ spent in $\ket{8}$.  As this was not accounted for by reducing the dwell time in $\ket{8}$, it results in an effective timing error.   In addition, there is a much smaller correction that is practically independent of the optical pulse time.  However, the influence of this term depends on the microwave and optical coupling strengths relative to the values determined by their respective $\pi$ times.  The uncertainties given in table~\ref{tab:system}  for the microwave and coupling errors corresponding to 1\% and 2.5\% uncertainty in the microwave and optical couplings, respectively. 

A differential gravitational shift of $1.31(15)\times 10^{-18}$ is included to account for a height difference of 1.2(1)\,cm between the ions.  All other shifts are expected to contribute negligibly to the frequency difference compared to the statistical uncertainty achieved in the comparisons.  A discussion of these omitted shifts is given in methods. 

\begin{figure}
\centering
\includegraphics[width=0.9 \columnwidth]{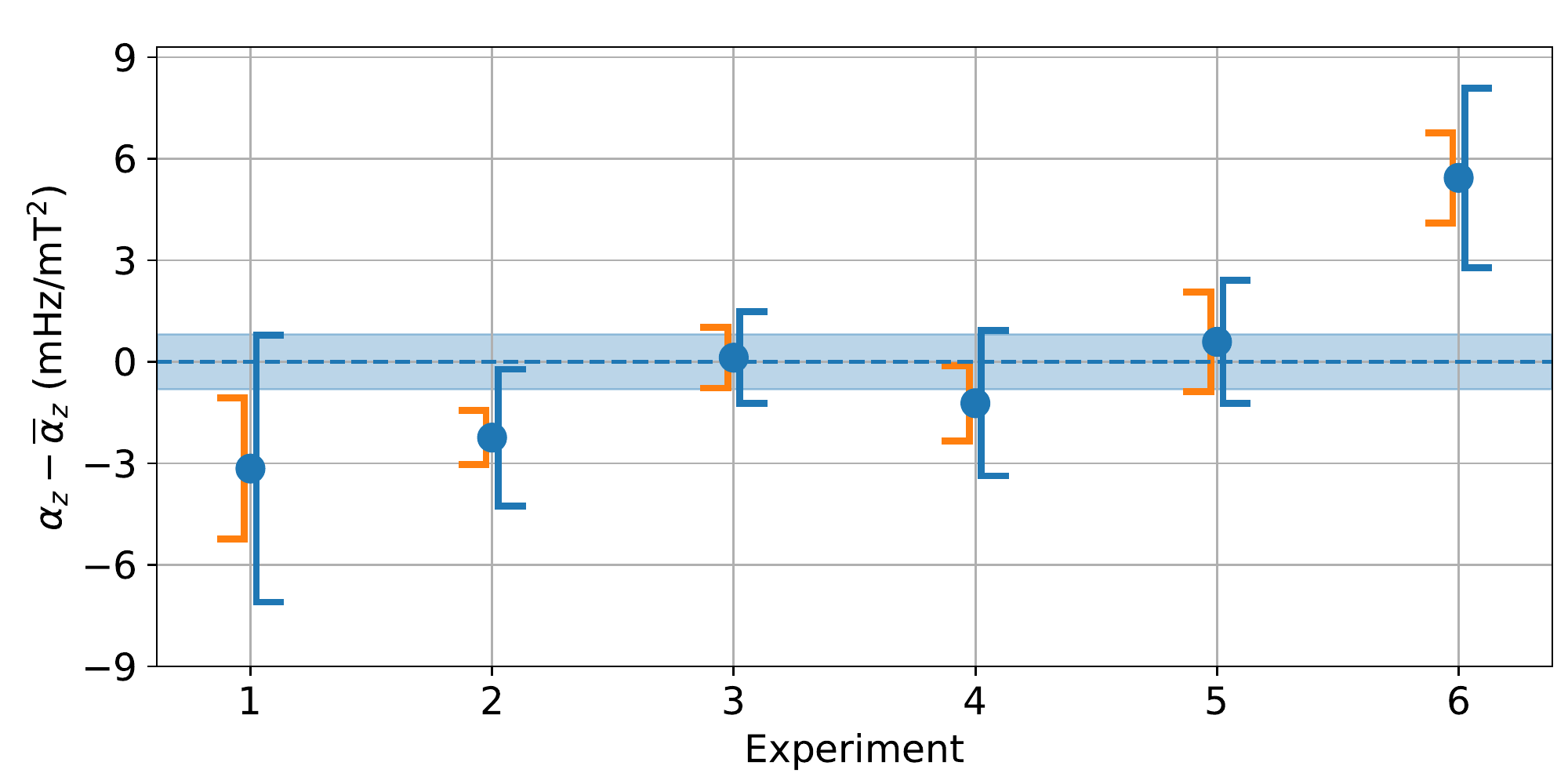}
\caption{ Measurements of the quadratic Zeeman coefficient ($\alpha_z$).  Uncertainty budgets and parameters $\tau_L, \tau_1, \tau_2,$ and $T$ and applied static magnetic fields $B_1$ and $B_2$ for each experiment are given in the supplemental. Each point is the mean of all data collected at the given parameters, corrected for systematic errors and offset by the weighted mean of all measurements.  Error bars on the left represent the statistical uncertainty and those on the right are inclusive of the systematic uncertainty.  The shaded region indicates the uncertainty in the weighted mean.  From all six measurements we obtain the estimate $\alpha_z=-4.89264(88)\,\mathrm{Hz/mT^2}$, with $\chi_\nu^2=1.30$. }
\label{Final_coeff}
\end{figure}

The quadratic Zeeman coefficient ($\alpha_z$) of the $\ket{^1S_0}\leftrightarrow\ket{^3D_1}$ transition can be determined by 
\begin{equation}
\alpha_z = \frac{\delta_{z}}{B_1^2-B_2^2},
\label{Eq:quadratic}
\end{equation}
where $B_k$ is the applied static magnetic field for Lu-$k$ and $\delta_{z}$ is the frequency difference between the two atomic references appropriately corrected for all other systematic shifts.  The results from six measurements are given in \fref{Final_coeff} offset by the weighted mean.  From all six measurements we obtain $\alpha_z=-4.89264(88)\,\mathrm{Hz/mT^2}$ with $\chi_\nu^2=1.30$ as discussed in methods.  With $B_1=1.9\,\mathrm{mT}$ and $B_2=0.1\,\mathrm{mT}$, the estimated uncertainty in $\alpha_z$ requires agreement of the two atomic references to within $9.0\times10^{-18}$ when held at the same magnetic field.  A further comparison (Expt. 7) was carried out with both systems at $0.1\,\mathrm{mT}$ to confirm this, which gave the result $(9.2\pm(3.6)_\mathrm{stat}\pm(9.1)_\mathrm{sys})\times10^{-18}$ as determined by 4 separate measurements with a total integration time of 29.3h.  As indicated in table~\ref{tab:system}, the comparison is limited by the assessment of the probe-induced ac-Stark shift. 

Probe-induced ac-Stark shifts can be heavily suppressed using hyper-Ramsey spectroscopy~\cite{yudin2010hyper,hobson2016modified,kuznetsov2019effect} and the principle also applies to correlation spectroscopy and HARS.  Following the work of Yudin, et al.~\cite{yudin2010hyper,hyperRamsey2012peik}, a hyper-Ramsey pulse can be included, as indicated in \fref{fig:scheme}c, with the clock laser frequency shifted by the measured ac-Stark shift during the clock pulses, and the additional hyper-Ramsey $\pi$-pulse phase shifted by $\pi$ with respect to the first $\pi/2$-pulse.  As shown in the supplemental, this not only suppresses the ac-Stark shift, but also eliminates the effective timing error from the finite optical $\pi$-time and the sensitivity to the optical coupling strength.

\begin{figure}
\centering
\includegraphics[width=0.9\columnwidth]{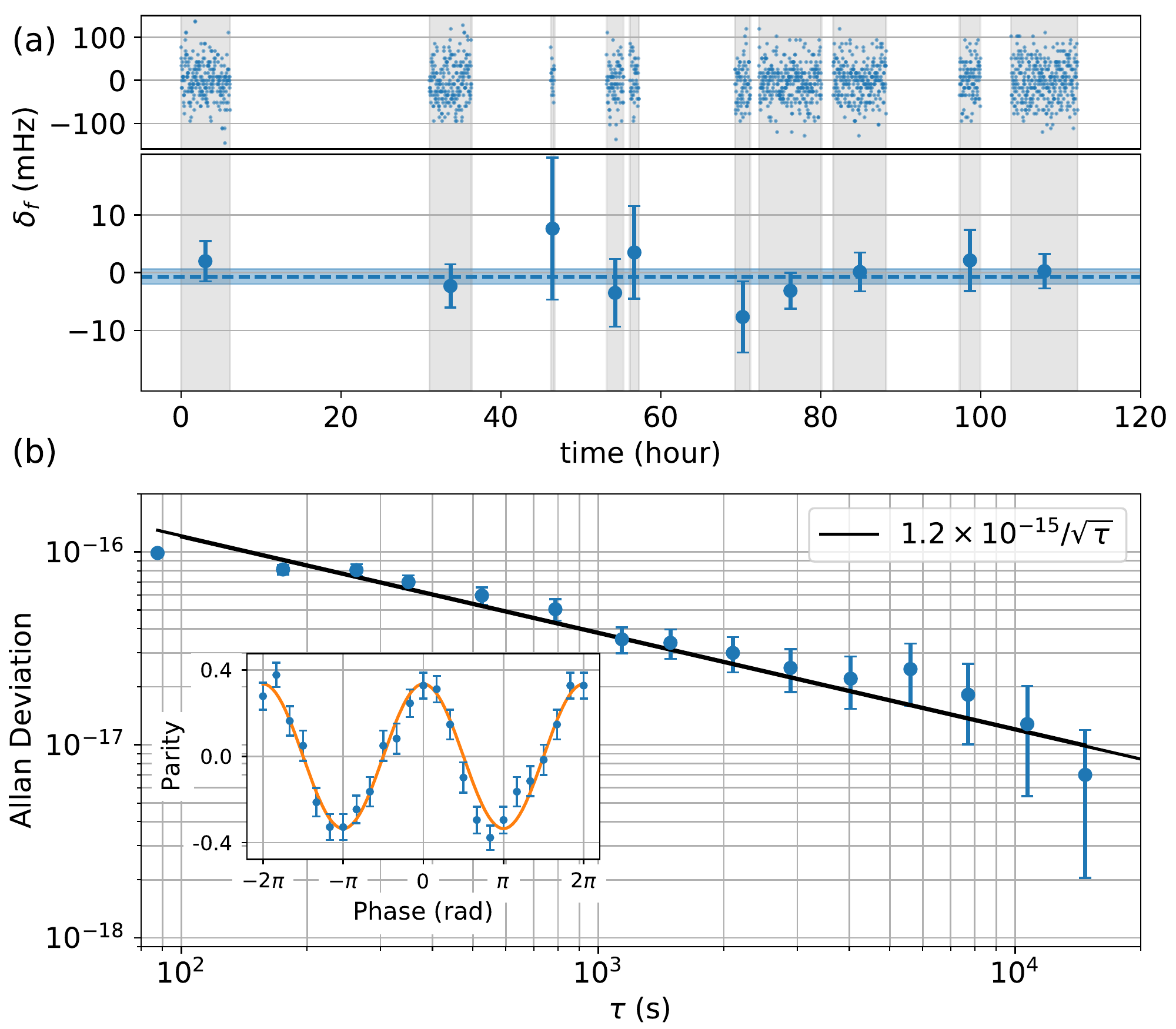}
\caption{{\bf Measurement instability.} Comparison data between two $^{176}$Lu$^+$ references implemented via HARS including hyper-Ramsey for ac-Stark shift suppression (Expt. 8). (a) The frequency difference measured over 42.4 hours in ten runs of the experiment.  Points are  the mean of each run and error bars are the statistical error determined by the projection noise limit.  The dashed horizontal line is the weighted mean ($\chi^2_\nu=0.45$) and the shaded region the corresponding uncertainty, which represents a fractional frequency difference of $(-2.0\pm(3.7)_\mathrm{stat}\pm(0.9)_\mathrm{sys}) \times 10^{-18}$. (b)  Fractional Allan deviation for the longest single run.  The inset shows a typical parity signal as a function of the relative phase of the second $\pi/2$-pulse, the amplitude of which is limited by heating in Lu-1.}
\label{compareHR_allan}
\end{figure}

A final comparison (Expt. 8) including a hyper-Ramsey pulse gave $(-2.0\pm(3.7)_\mathrm{stat}\pm(0.9)_\mathrm{sys}) \times 10^{-18}$ as determined by ten measurements with a total integration time of $42.4\,\mathrm{h}$.  Data is shown in \fref{compareHR_allan} and an error budget given in table~\ref{tab:systemexp8}.  With suppression of the probe-induced ac-Stark shift, the only shifts above $10^{-18}$ are the SD shift and the gravitational shift and these are still less than the measurement precision.  Moreover, the SD shift is primarily due to a heating problem in one trap.  Consequently, clock agreement at this level is not reliant on substantial corrections and control of large systematics as it is for all other clock comparisons that have been made thus far.
\begin{table*}[ht]
\caption{Uncertainty budget for comparison using Hyper-Ramsey (Experiment 8) for which $\tau_L=8\,\mathrm{ms}, \tau_1=\tau_2=24\,\mathrm{ms},$ and $T=200\,\mathrm{ms}$ with an applied static magnetic field of $\approx 0.1\,\mathrm{mT}$ for both traps. All values are relative to $10^{-18}$ of the HA frequency.}
\label{tab:systemexp8}
\begin{tabular*}{\textwidth}{@{\extracolsep{\fill}} >{\quad}l r r r r r r <{\quad}}
\toprule[0.75pt]
 & \multicolumn{2}{c}{Lu-1} &  \multicolumn{2}{c}{Lu-2} & \multicolumn{2}{c}{Difference} \\
\cmidrule[0.5pt](lr){2-3} \cmidrule[0.5pt](lr){4-5} \cmidrule[0.5pt](r{0.75em}){6-7}
Effect & \multicolumn{1}{c}{Shift} & \multicolumn{1}{c}{Unc.} & \multicolumn{1}{c}{Shift} & \multicolumn{1}{c}{Unc.} & \multicolumn{1}{c}{Shift} & \multicolumn{1}{r<{\quad}}{Unc.} \\  \hline
Excess micromotion &-0.41&0.37 &-0.44 &0.34&0.03 &0.50\\
Second-order Doppler (thermal)&-1.87&0.45&-0.13&0.06 &-1.75&0.45\\
ac-Zeeman (rf)&0.54&0.01&0.15&0.01&0.39 &0.01 \\
ac-Zeeman (microwave)&-0.06&0.03 &-0.13&0.11&0.07& 0.11\\
Gravity shift\footnote{Only differential shifts are considered.}&-&-&-&-&-1.31&0.15 \\
Microwave coupling&0&0.21 &0&0.21&0& 0.30\\
Residual quadruple shift&0.22&0.02&0&0.32&0.22& 0.32\\
\textbf{Total shift} &-1.58&0.68&-0.55&0.60&-2.34 &0.93\\
\bottomrule[0.75pt]
\end{tabular*}
\end{table*}

In summary, we have used correlation spectroscopy to compare two $^{176}$Lu$^+$ atomic references with an inaccuracy at the low $10^{-18}$ level.  This has provided a high accuracy assessment of the quadratic Zeeman shift, which is the leading systematic for the $^{176}$Lu$^+$ $^1S_0\leftrightarrow{^3D_1}$ clock transition.  Hyper-Ramsey spectroscopy is also compatible with HARS and practically eliminates the probe-induced ac-Stark shift as the next leading systematic.  This then provides a frequency reference with systematics that are easily controlled to an inaccuracy below $10^{-18}$ without the technical challenges faced by other systems.

An accuracy claim for an individual system would require an assessment of the blackbody radiation (BBR) shift.  Presumably the use of simulations and thermal cameras\cite{YbPeik,dolevzal2015analysis} could equally be used here, although it is unclear how to quantify such an assessment.  A crude analysis as given in the methods would bound the temperature to $35(10)^\circ\mathrm{C}$, which, due to the low BBR shift for the $^1S_0\leftrightarrow{}^3D_1$ transition, corresponds to an BBR shift of just $-1.56(27)\times 10^{-18}$ and gives a total uncertainty for both systems of $<8\times10^{-19}$.  However, although it is done elsewhere \cite{YbPeik, brewer2019al+}, we refrain from making any accuracy claims without a same-species comparison to the appropriate level.  Here stability limits us to the low $10^{-18}$, which is primarily limited by a heating problem in Lu-1.

With the lower heating rate and real-time monitoring of EMM, systematics at the mid $10^{-19}$ are readily achievable.  Correlation spectroscopy with a 10\,s interrogation time was demonstrated in our previous work~\cite{tan2019suppressing} albeit within the same trap.  In separate chambers, uncorrelated magnetic field noise may be a limiting factor.  However, at the operating field of $0.1\,\mathrm{mT}$, the maximum magnetic sensitivity of the 848-nm clock states is $\sim 0.4\,\mathrm{Hz/\mu T}$.  This is two orders of magnitude smaller than Al$^+$, for example, for which lifetime-limited correlation spectroscopy has been claimed \cite{clements2020lifetime}.  Increasing the interrogation time to an anticipated 10\,s would make comparisons below $10^{-18}$ achievable on the time scale of a few days.

The relative ease at which a comparison at this level can be achieved with this system should not be under-rated.  Comparisons between clock systems claiming this level of accuracy \cite{collaboration2021frequency, dorscher2021optical} have variations of the measured ratios that cannot be explained by the error budgets of the individual systems, which is compelling evidence that large systematic shifts are not being controlled to the levels claimed.  Such problems are unlikely to occur for systems such as lutetium or thullium \cite{thullium2021} for the simple fact that systematics are not large enough to cause such variations.  Indeed, the total shift and any single contribution given in table~\ref{tab:systemexp8} is, at most, at the level of the current measurement precision.  Moreover, it is self-evident that systems having such low systematics have substantial room for improvement if current state-of-the-art technologies were employed to take advantage of this potential.  This is particularly noteworthy for lutetium as the magnetic field sensitivity of the clock states are the lowest of any atomic system that we are aware of and the lifetime is practically indefinite.  Consequently, interrogation times will not be limited by lutetium in any foreseeable future. 

\noindent{\bf{{Acknowledgements}}}\\
This work is supported by: the National Research Foundation, Prime Ministers Office, Singapore and the Ministry of Education, Singapore under the Research Centres of Excellence programme; and the National Research Foundation, Singapore, hosted by National University of Singapore under the Quantum Engineering Programme (Grant Award QEP-P5).
\bibliographystyle{unsrt}
\bibliography{biblio2}
\clearpage


\section*{Supplementary material}

# Supplemental Material for $^{176}$Lu$^+$ clock comparison at the $10^{-18}$ level via correlation spectroscopy

Zhang Zhiqiang,$^1$ K. J. Arnold,$^{1,2}$ R. Kaewuam,$^1$ and M. D. Barrett$^{1,3,4,*}$

$^1$*Centre for Quantum Technologies, 3 Science Drive 2, 117543 Singapore*
$^2$*Temasek Laboratories, National University of Singapore, 5A Engineering Drive 1, 117411 Singapore*
$^3$*Department of Physics, National University of Singapore, 2 Science Drive 3, 117551 Singapore*
$^4$*National Metrology Centre, Agency for Science,
Technology and Research (A\*STAR), 8 CleanTech Loop, Singapore 637145*

We provide addition information relevant to the $^{176}$Lu$^+$ clock comparison. A discussion of additional shifts that have been considered but do not affect the comparisons reported are given. We evaluate the sensitivity of hyperfine averaged Ramsey spectroscopy (HARS) to imperfections in timing, optical coupling, and microwave couplings. In relation to this we include a discussion of the Physikalisch-Technische Bundesanstalt (PTB) demonstration of hyper-Ramsey spectroscopy in the context of what it means for our system. We also include: expressions for evaluating the microwave ac-Zeeman systematic; a back-of-the-envelop estimate for a reasonable temperature rise for our systems; and systematic uncertainty budgets for Experiments 1 to 6.

## I. ADDITIONAL CLOCK SHIFTS.

Clock shifts considered but not included in the error budget are given here. These shifts do not affect the comparisons made in this work under any realistic circumstances.

**AC-Zeeman shifts:** Additional ac-Zeeman shifts arise from two sources. The first concerns the ac coupling between Zeeman states. These do not cause a clock shift directly, but they do modify the effective $g$-factor of the level or equivalently the B-field determined by measuring a Zeeman splitting. With the rf field at the $\mu$T level, this only affects the measured field at the $10^{-9}$ level. At 50 Hz, we estimate field noise to be around 8 nT rms, which similarly affects the measured field at the $10^{-9}$ level. The second source arises from the coupling between fine-structure levels. This has both a scalar and tensor contribution, but the latter is eliminated by hyperfine averaging (HA). The former is determined by $\alpha_z \langle B^2 \rangle$ where $B$ is the total ac-field. This has the same magnitude as the measured perpendicular component, as determined by measurements from three orthogonal directions [1]. For determination of $\alpha_z$, this influences $\langle B_1^2 \rangle - \langle B_2^2 \rangle$ by $\lesssim 10^{-6}$ fractionally, compared to the $1.7 \times 10^{-4}$ fractional inaccuracy of $\alpha_z$. Similarly, the influence on a comparison when the clocks are both at low field is $\sim 10^{-20}$.

**The blackbody radiation (BBR) shift:** The polarizability for the 848-nm transition in $^{176}$Lu$^+$ has been very well characterized [2, 3]. The corresponding BBR shift is the lowest of any clock candidate under active investigation with an estimated shift of $-1.36(10) \times 10^{-18}$ at 300 K. For the comparison result reported we are only concerned with the frequency difference. Even allowing for one trap at 25 °C and the other at 45 °C, the contribution to the frequency difference would be just $\sim 5 \times 10^{-19}$, and well below our measurement resolution. Given the near identical construction of the two traps and the similar trap drive powers supplied to each, such a temperature differential is highly unlikely. We also note that a temperature assessment with an inaccuracy of 2 °C has been reported [4], which was based on measurements made on a trap of identical construction, with simulations to account for different mounting within the vacuum chambers [5]. In our case the mounting is also identical. In such an assessment it must be assumed that traps of near identical construction have near identical temperatures, at least better than 2 °C in fact. Although we would not be so bold as to suggest the traps have a temperature difference of $\lesssim 2$ °C, we think it reasonable to assume the temperature difference is significantly less than 20 °C and not have any relevance to the comparisons made here. As we are unable to make a measurement of the temperature in either system and any reasonable temperature difference is immaterial to the final result, we do not include this in the comparison assessment. In addition, we give a back-of-the envelope analysis in section IV to crudely bound the temperature and hence BBR shift for each system.

**AOM Chirp:** The switching of acousto-optic modulators (AOMs) and microwave sources can potentially introduce artificial frequency shifts [7] even when path lengths are nominally stabilized. Following the methodology reported in Ref.[8], this was investigated in preliminary experiments and estimated to be below $10^{-19}$ for the experiments reported here.

**Probe induced ac-Stark shift:** For the final comparison, which includes a hyper-Ramsey pulse, we have omitted any contribution from the probe-induced ac-Stark shift. Similar to experiment 7, the probe-induced ac-Stark shift was approximately 7 Hz during the pulses with a differential between the two-systems of 0.52(0.47) Hz. As evidenced by experiment 7, the effect on the comparison is already at the level of the measurement precision, even without the hyper-Ramsey pulse. Conse-

---

* phybmd@nus.edu.sg

quently, any level of suppression would be beyond our measurement precision. As shown in section II, for small microwave detunings, the shift has the same cubic dependence on the effective ac-Stark shift with only minor modifications from the microwave detunings. Hence, even if the frequency was not shifted during the optical pulses, the shift would only influence the comparison at the mid $10^{-19}$ level. As the net effective shift is trivially stable at the 0.5 Hz level, the shift of each system will be limited by the microwave coupling accuracy, which is already accounted for in the error budgets.

**Timing accuracy:** For HARS, timing accuracy of the pulse sequence can influence the dwell time in each state, which could potentially degrade the averaging. If the dwell time for each $F$ is $T + \tau_1 + \tau_2 + \delta T_F$, then, using Eq. 7 from Ref[6], the clock shift $\delta$ will be

$$\delta = \frac{1}{T_R} \sum_F \Delta_F \delta T_F = \frac{\Delta_2 \delta T_6 - \Delta_1 \delta T_8}{T_R}$$

where $\Delta_F$ are linear combinations of the microwave detunings $\Delta_1, \Delta_2$ as defined in Ref[6]. For the second equality we have used the fact that $\langle \delta T_F \rangle = 0$ as any non-zero value can be absorbed in the definition of $T$. The on-board 100 MHz clocks of the digital I/O boards are externally synchronized so the main considerations are switching delays of clock laser AOM's and microwave sources. Switching delays of the clock laser AOM's are $\sim 60$ ns. However, as this doesn't influence the overall dwell time in $|8\rangle$, it doesn't result in a clock shift. Switching time differences of microwave sources only influence the sychronization at the 5 ns level. Hence, for the Hertz level detunings of the microwave fields, fractional shifts are well below $10^{-21}$ for all experiments reported here.

**Imperfections in the implementation of HARS:** As noted in the main text, microwave detunings, the finite duration of the optical pulses, and imperfect coupling strengths of the microwave and optical fields can compromise the effectiveness of our interrogation technique. A detailed analysis of these effects are provided in section II, but we summarize the main findings here. The leading order effect is due to the finite duration of the optical pulses, which is associated with an effective time $2\tau_L/\pi$ spent in $|8\rangle$. When not accounted for, it results in an effective timing error given by Eq. S3. An additional error given by Eq. S17 has the same form, which arises from deviations of the optical coupling $\Omega_L$ from the ideal condition $\Omega_L \tau_L = \pi$. From the analysis following Eq. S19, both of these corrections are practically eliminated when including a hyper-Ramsey pulse. As seen in Eq. S13, there is an additional term that is practically independent of $\tau_L$, and this term is very well approximated by the analytic expression given in Eq. S15, which is an oscillatory function with an amplitude that is quadratic in the microwave detunings normalized by the coupling strengths. For the microwave detunings used here, this term has no significant contribution to the analysis and is only slightly modified when including a hyper-Ramsey pulse. However, a contribution from this term arises from imperfect microwave couplings as discussed in relation to Eq. S18. As illustrated in Fig. 5 and Fig. 6 of the supplemental, the effect is to add a constant to the quadratic amplitude, and we use this amplitude to bound the contribution as an uncertainty.

Microwave detunings used in this work were measured and set with a resolution of approximately 0.1 Hz. However, at the larger operating field of 1.9 mT, variations in the magnetic field could change the detunings by up to $\pm 2$ Hz. Data analysis has included these variations using the corrections provided by Eq. S3. As a consistency check, Eq. S16 was also used to confirm that it gave no statistically significant difference in the final results.

**Uncertainty in $\boldsymbol{\alpha_z}$**: Complete measurement results and uncertainty budgets for experiments 1-6 are given in section V. The leading systematic is due to the ac-Stark shift, which is uncorrelated as it is remeasured between experiments. For this reason the statistical and ac-Stark uncertainties are summed in quadrature and only these used to evaluate the weighted mean value $\alpha_z = -4.89264(81)$ with $\chi_\nu = 1.3$. The HARS timing systematic is correlated across experiments as all microwave detunings are dependent on just a few parameters via (S3) and (S4). These parameters are randomly resampled within their stated uncertainties to find the additional variance in the weighted mean $\alpha_z$. The standard deviation is added in quadrature to the uncertainty in $\alpha_z$ above to arrive at $-4.89264(88)$. All other systematics are generally correlated across experiments but do not contribute significantly to the uncertainty.

## II. HARS SENSITIVITY FACTORS

The Hamiltonian governing hyperfine averaged Ramsey spectroscopy (HARS) is given by [6]

$$H = \delta |g\rangle \langle g| - \sum_F \Delta_F |F\rangle \langle F| + \frac{\Omega_L}{2} \left( |\bar{F}\rangle \langle g| + |g\rangle \langle \bar{F}| \right)$$
$$+ \frac{\Omega_2}{2} \left( |6\rangle \langle 7| + |7\rangle \langle 6| \right) + \frac{\Omega_1}{2} \left( |7\rangle \langle 8| + |8\rangle \langle 7| \right). \quad \text{(S1)}$$

where detunings

$$\Delta_6 = \frac{2\Delta_2 + \Delta_1}{3}, \quad \Delta_7 = -\frac{\Delta_2 - \Delta_1}{3}, \quad \text{(S2a)}$$

and

$$\Delta_8 = -\frac{2\Delta_1 + \Delta_2}{3}, \quad \text{(S2b)}$$

are linear combinations of the microwave detunings $\Delta_1$ and $\Delta_2$ from the $|8\rangle \leftrightarrow |7\rangle$ and $|7\rangle \leftrightarrow |6\rangle$ transitions respectively.

In [6] we demonstrated the use of this technique in reducing inhomogeneous broadening. However the result

3

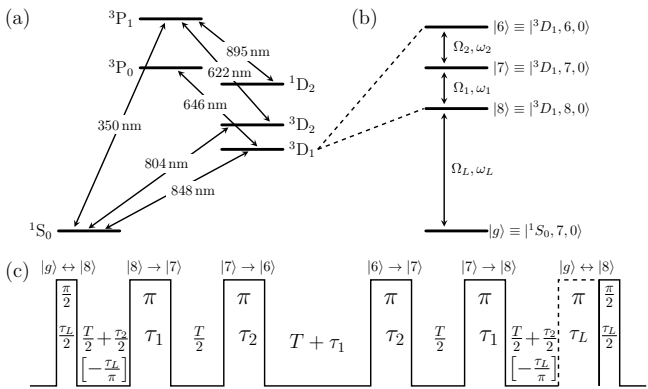

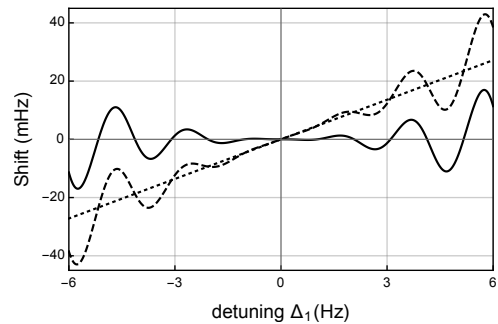

FIG. 1. Timing sequence followed for the simulations presented here. The additional $\pi$ pulse (dashed) is optionally included for hyper-Ramsey spectroscopy. Reduction of the dwell time in $|8\rangle$ by $\frac{2\tau_L}{\pi}$ (terms in square brackets) can compensate for the time spent in $|8\rangle$ during the optical $\pi/2$ pulses when not using hyper-Ramsey.

was derived in the limit of small detunings and neglected the finite duration of the enclosing optical Ramsey pulses. Given the levels of accuracy we are now achieving it is pertinent to reconsider those limits. Here we simulate the full equations to illustrate the effects of operating outside these limitations and compare with analytic calculations. We use the timing sequence shown in Fig. 1 which was used in our recent comparison work as a case study. Although the frequency shifts can be numerically calculated for a given set of parameters, we provide a number of analytic approximations that accurately determine the shifts and illustrate how they are influenced by the various experimental parameters. Consequently it is sufficient to take $\tau_1 = \tau_2 = \tau = 24$ ms, $T = 0.2$ s, and $\tau_L = 8$ ms as default values for the purpose of illustration.

The main limitation that the optical $\pi$-time $\tau_L$ imposes is that it does not account for the effective time spent in the upper state during the pulse. This can be accounted for by reducing the dwell time in $\left|\bar{F}\right\rangle = |8\rangle$ by $2\tau_L/\pi$, which we henceforth refer to as the compensated sequence. The $2/\pi$ scale factor can be understood in the context of the sensitivity factor introduced in [9]. As this can be considered a timing error, we expect a frequency error given by

$$\delta = -\frac{\frac{2}{\pi}\tau_L}{3(T+2\tau) + \frac{2}{\pi}\tau_L}\Delta_8, \quad \text{(S3)}$$

which is easily derived from the considerations given in [6]. However, this does not take into account any further deviations arising from the microwave detunings. To illustrate this, we plot the actual shift as a function of $\Delta_1$ in Fig. 2, both with and without the compensation time, and the linear approximation given by Eq. (S3). As seen in the figure, the linear approximation captures the overall trend but there are further corrections that should be considered at our current level of accuracy. In addition,

FIG. 2. Shift in the clock frequency as a function of the microwave detuning $\Delta_1$ with (solid line) and without (dashed) compensating for the optical $\pi$-time. The dotted straight line is the linear approximation given by Eq. (S3). Parameters are given by $\tau_1 = \tau_2 = \tau = 24$ ms, $T = 0.2$ s, and $\tau_L = 8$ ms.

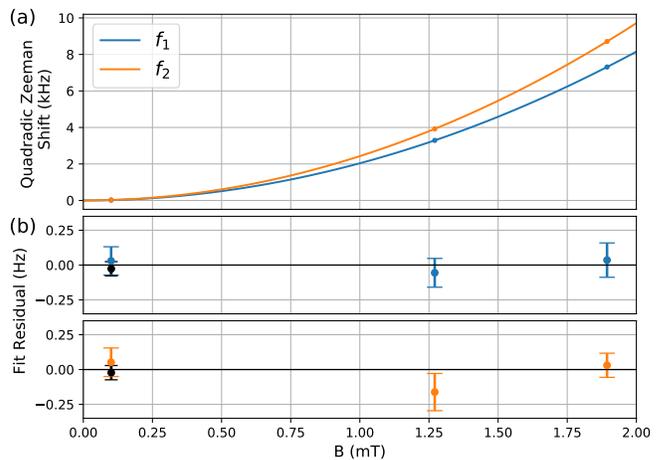

FIG. 3. Calibration of the resonant microwave frequencies. (a) Each point is the weighted mean of all measurements in Lu-1 at the three operating magnetic fields. Solid lines are the quadratic fit. Frequencies are plotted as the shift relative to $f_{1,0}$ ($f_{2,0}$) (b) Residuals of the points with respect to the fit. Black points are the measurements in Lu-2.

it is also pertinent to consider the effect of imperfect coupling strengths of both the microwave and optical fields, which can also influence the accuracy achievable by the interrogation technique.

The resonant microwave frequencies $f_1$ and $f_2$ were measured at several points between experiments when resetting the microwave detunings $\Delta_k$ near zero. These were measured using Ramsey spectroscopy with between 50 ms and 200 ms interrogation times and had typical uncertainty of $\approx$0.1 Hz per measurement. From multiple measurements in Lu-1 at B = (0.1, 1.2, and 1.9) mT, we fit to the model

$$f_1 = f_{1,0} + \alpha_{78}B^2 \quad \text{(S4)}$$
$$f_2 = f_{2,0} + \alpha_{67}B^2 \quad \text{(S5)}$$

as shown in Fig. 3 and find parameters

$$f_{1,0} = 10\,491\,519\,945.035(92)\,\text{Hz}$$
$$f_{2,0} = 11\,290\,004\,290.255(95)\,\text{Hz}$$
$$\alpha_{78} = 2\,038.533(44)\,\text{Hz/mT}^2$$
$$\alpha_{67} = 2\,428.945(37)\,\text{Hz/mT}^2.$$

Lu-2 was only operated at 0.1 mT but microwave measurements are in agreement with this model.

The magnetic field was determined with a projection-noise-limited uncertainty of $\approx 20\,\text{nT}$ per measurement by measuring the Zeeman splitting between $|6, \pm 1\rangle$ and using the previously determined $g$-factors [10]. Measurements were interleaved with each clock interrogation and the field was observed to drift by no more than $\pm 200\,\text{nT}$ throughout the course of any experiment. During each experiment the microwave drive frequencies $\omega_k$ were fixed and the detunings $\Delta_k$ were evaluated in later analysis from the interleaved magnetic field measurements and Eqs. (S4,S5). For all experiment $|\Delta_k| < 2\,\text{Hz}$. The uncertainties in the average $\Delta_k$ for each experiment, and thus the resulting HARS timing error systematic, are limited by the precision of $f_k$ from Eqs. (S4,S5).

Coupling strengths, or equivalently the $\pi$-times, for both microwave and optical transitions were determined from Rabi flopping curves on the respective transition, with a typical uncertainty $\sim 1\%$. Drifts in the optical coupling can be bounded to $\pm 2.5\%$ from the stability of repeated measurements of the ac-Stark shift. For the microwave couplings, repeated measurements throughout the measurement campaign indicate a $\pm 1.0\%$ stability.

To derive an expression for the error we first note that an error signal is derived from the HARS sequence by measuring the population difference between two sequences in which the last pulse is phase shifted by $\pm \pi/2$. The frequency error is then determined by the detuning $\delta$ for which

$$|c_+|^2 - |c_-|^2 = 4\,\text{Re}(c_1 c_2^*) = 0 \quad \text{(S6)}$$

where $c_\pm$ are the final population amplitudes for the sequence with a corresponding phase shift of $\pm \pi/2$ and $c_\pm = c_1 \pm c_2$ are introduced for convenience. Deriving expressions for $c_1$ and $c_2$ is facilitated by expressing the operators for the various pulses in terms of the functions

$$f(x,y,a) = \cos\left(a\sqrt{x^2+y^2}\right) + i\frac{y}{\sqrt{x^2+y^2}}\sin\left(a\sqrt{x^2+y^2}\right), \quad \text{(S7)}$$

and

$$g(x,y,a) = \frac{x}{\sqrt{x^2+y^2}}\sin\left(a\sqrt{x^2+y^2}\right).$$

Equations for the frequency error for both the compensated and uncompensated sequences are almost identical and may be written

$$\text{Re}\left\{ie^{i\theta}F(\bar{\Omega}_L, X)G(\bar{\Omega}_1, \bar{\Omega}_2, \bar{\Delta}_1, \bar{\Delta}_2)\right\} = 0$$

where

$$F(\bar{\Omega}_L, X) = 4f^2(\bar{\Omega}_L, X, \tfrac{\pi}{4})g^2(\bar{\Omega}_L, X, \tfrac{\pi}{4})e^{-2iX} \quad \text{(S8)}$$

and

$$G(\bar{\Omega}_1, \bar{\Omega}_2, \bar{\Delta}_1, \bar{\Delta}_2) = f^2\left(\bar{\Omega}_1, -\bar{\Delta}_1, \tfrac{\pi}{2}\right)e^{i\pi\bar{\Delta}_1/2}e^{-i\phi_1} - g^2\left(\bar{\Omega}_1, \bar{\Delta}_1, \tfrac{\pi}{2}\right)f^2\left(\bar{\Omega}_2, -\bar{\Delta}_2, \tfrac{\pi}{2}\right)e^{i\pi\bar{\Delta}_2/2}e^{-i\phi_2} + g^2\left(\bar{\Omega}_1, \bar{\Delta}_1, \tfrac{\pi}{2}\right)g^2\left(\bar{\Omega}_2, \bar{\Delta}_2, \tfrac{\pi}{2}\right). \quad \text{(S9)}$$

In the above expressions

$$\bar{\Omega}_k = \Omega_k \tau/\pi, \quad \bar{\Omega}_L = \Omega_L \tau_L/\pi, \quad \bar{\Delta}_k = \Delta_k \tau/\pi,$$

$$X = (\delta + \Delta_8)\tau_L/\pi,$$

$$\phi_1 = (T+2\tau)(2\Delta_1 + \Delta_2) - \Delta_1 \tau/2,$$

and

$$\phi_2 = \left(T + \tfrac{3}{2}\tau\right)\Delta_2.$$

The only difference in the equations, for the compensated and uncompensated case is in the parameter $\theta$, which is given by

$$\theta = \delta\left[3(T+2\tau)\right] = \delta T_R, \quad \text{(S10)}$$

in the compensated case or

$$\theta = \delta\left[3(T+2\tau) + 2\tau_L/\pi\right] + 2\Delta_8 \tau_L/\pi,$$
$$= \delta T_R + \tfrac{2\tau_L}{\pi}\Delta_8, \quad \text{(S11)}$$

in the uncompensated case. In both cases $T_R$ is the effective Ramsey time for the sequence, which differs only by the compensation time $2\tau_L/\pi$.

Solution to Eq.(S6) is best understood by first considering the case in which $\bar{\Omega}_L = \bar{\Omega}_k = 1$ and using the approximation $F(1, X) \approx 1$. In a series expansion of $F(1, X)$, odd and even terms are purely imaginary and real respectively, with the next non-zero order being $\approx 0.1 i X^3$. Consequently the approximation is exceptionally good under practical operating conditions and leads

to the expression

$$\text{Re}\left\{ie^{i\theta}G(\bar{\Delta}_1, \bar{\Delta}_2)\right\} = 0 \quad (S12)$$

where we have dropped the parameters $\bar{\Omega}_k$ for notational convenience. Thus,

$$\delta = -\frac{\frac{2}{\pi}\tau_L}{T_R}\Delta_8 - \frac{1}{T_R}\tan^{-1}\left(\frac{\text{Im}\left[G(\bar{\Delta}_1, \bar{\Delta}_2)\right]}{\text{Re}\left[G(\bar{\Delta}_1, \bar{\Delta}_2)\right]}\right), \quad (S13)$$

where the first term only applies for the uncompensated case and is exactly the timing error given in Eq. (S3). The second term is common to both and only differs in the definition of $T_R$, which has the only dependence on $\tau_L$ for the second term. This illustrates the separation of the error into a timing error associated with the uncompensated finite time of the optical pulse and that associated with the microwave detunings.

A further simplification makes use of the approximation

$$f^2\left(1, -x, \tfrac{\pi}{2}\right) e^{i\pi x/2} \approx -x^2. \quad (S14)$$

As is the case for $F(1, X)$, odd and even terms in the series expansion are purely imaginary and real respectively. Moreover, the third order term vanishes. Thus the imaginary component only appears at fifth order in $x$. As $g^2(1, x, \pi/2)$ is strictly real and even, the same conclusion applies to the second term in Eq. S9. This leads to the approximation

$$\delta = -\frac{\frac{2}{\pi}\tau_L}{T_R}\Delta_8 + \frac{\bar{\Delta}_2^2 \sin(\phi_2) - \bar{\Delta}_1^2 \sin(\phi_1)}{T_R}, \quad (S15)$$

which also utilizes the approximation $\tan^{-1} x \approx x$ consistent with the second order expansion given in Eq. (S14). This expression accurately captures the oscillatory behaviour of the correction relative to Eq. S3 and its dependence on $\Delta_k$, $\tau$ and $T$.

The quality of the approximations given in Eq. (S13) and Eq. (S15) is illustrated in Fig. 4, which shows the difference in the estimated clock shift between the value found numerically without any approximations and those found using Eq. (S15) (black) and Eq. (S13) (blue) with the latter increased by a factor of 100 to better demonstrate the dependence on detuning. Over the practical range of $\pm 2\,\text{Hz}$ the approximation from Eq. (S15) is better than $5 \times 10^{-20}$ of the optical frequency and Eq. (S13) is practically exact. For the parameters chosen, the $\pm 2\,\text{Hz}$ corresponds to about 10% of the microwave Rabi frequency.

The effect of non-ideal couplings for the optical and microwave fields can be determined by setting $\bar{\Omega}_L = 1 + q$ and $\bar{\Omega}_k = 1 + q_k$ and including $q$ and $q_k$ in the relevant expansion. For the optical field it can be shown that

$$F(1+q, X) \approx \cos^2\left(\tfrac{\pi}{2}q\right) e^{2ih(q)X} + \mathcal{O}(qX^2) \quad (S16)$$

where

$$h(q) = \tfrac{1}{1+q}\tan\left(\tfrac{\pi}{2}q\right) + \tfrac{1}{1+q}\sec\left(\tfrac{\pi}{2}q\right) - 1$$
$$\approx \left(\tfrac{\pi}{2} - 1\right)q.$$

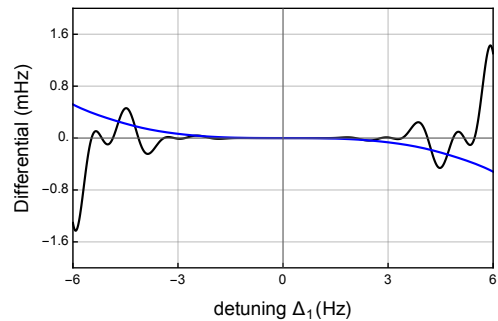

FIG. 4. Plots show the difference in the estimated clock shift between the value found numerically without any approximations and those found using Eq. (S15) (black) and Eq. (S13) (blue) with the latter increased by a factor of 100 to better highlight the dependence on detuning. Parameters are as for Fig. 2.

The amplitude factor has no influence on the resulting equation for $\theta$ but the phase factor results in an effective timing error

$$-\frac{\frac{2}{\pi}\tau_L}{T_R}\Delta_8 h(q) \approx -\frac{\frac{2}{\pi}\tau_L}{T_R}\Delta_8\left(\tfrac{\pi}{2} - 1\right)q \quad (S17)$$

and will be present in either the compensated or uncompensated case. For the experiments reported here this would be $\sim 10^{-17}q$. For $q < 0.1$, this is $\lesssim 10^{-18}$, so this effect would not have a significant effect on any results stated in the report. Furthermore, as the ac-Stark shift was measured at several points over the course the measurements, we can conclude that the coupling $\Omega_L$ was stable to at least the $\sim 2.5\%$ level. Moreover, we would expect any variations to average down, so any bias of a measurement would arise from the initial assessment of the $\pi$-time which effectively determines $\Omega_L$. We would expect it to be as probable for the $\pi$-time to be over-estimated as it is to be underestimated and hence this should not bias the average over multiple measurements. Hence this effect would not affect our results at the level of precision claimed.

For the microwave fields, we follow a similar procedure. Writing $\bar{\Omega}_1 = 1 + q_1$, we obtain the second order expansion

$$f^2\left(1+q_1, -x, \tfrac{\pi}{2}\right) e^{i\pi x/2} \approx -(x - i\tfrac{\pi}{2}q_1)^2,$$

and this also applies when including the additional $g^2$ function that appears in the second term of Eq. S9. The effect of imperfect coupling then is to modify the phases $\phi_k$ in Eq. S15 and replace the amplitudes $\bar{\Delta}_k^2$ by $\bar{\Delta}_k^2 + (\pi q_k/2)^2$. Hence the quadratic envelope of the oscillations is now offset by $(\pi q_k/2)^2$ as illustrated in Fig. 5. For $\bar{\Delta}_k \ll 1$, the modified error curve can be approxi-

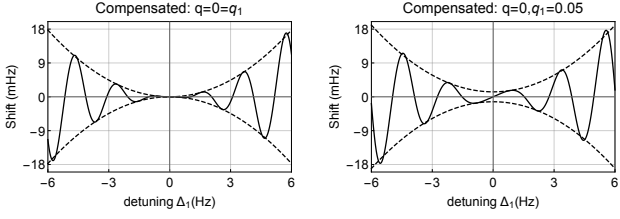

FIG. 5. Shift in the clock frequency as a function of the microwave detuning $\Delta_1$ compensated for the optical $\pi$-time. Plot on the left assumes ideal microwave and optical couplings and the dashed lines show the expected quadratic envelope as determined by Eq. S15. Plot on the right has the microwave coupling $\Omega_1$ increased by 5% ($q_1 = 0.05$) and the dashed lines are the offset quadratic envelope as discussed in the text. Parameters for both are as for Fig. 2.

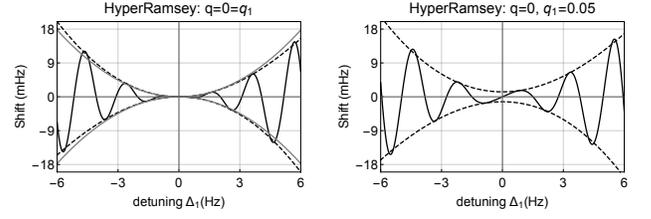

FIG. 6. Shift in the clock frequency as a function of the microwave detuning $\Delta_1$ for the hyper-Ramsey sequence. Plot on the left assumes ideal microwave and optical couplings and the dashed lines show the expected quadratic envelope modified by the additional term in Eq. S19. For comparison, the gray lines give the quadratic envelope for the compensated sequence. Plot on the right has the microwave coupling $\Omega_1$ increased by 5% ($q_1 = 0.05$) and the dashed lines are the offset envelope as discussed in the text. Parameters for both are as for Fig. 2.

mated by adding

$$\delta_m = \frac{\left(\bar{\Delta}_2^2 - \left(\frac{\pi q_2}{2}\right)^2\right) \sin \phi_2 + \pi \bar{\Delta}_2 q_2 \cos \phi_2}{T_R} - \frac{\left(\bar{\Delta}_1^2 - \left(\frac{\pi q_1}{2}\right)^2\right) \sin \phi_1 + \pi \bar{\Delta}_1 q_1 \cos \phi_1}{T_R}, \quad (S18)$$

to Eq. S15.

An analytic result for the inclusion of a hyper-Ramsey pulse is significantly more tedious, but the same procedure can be followed. This leads to a similar equation with the form

$$\text{Re}\left\{ie^{i\theta}\bar{F}(X)G(\bar{\Delta}_1, \bar{\Delta}_2) + \bar{H}(X, \bar{\Delta}_1, \bar{\Delta}_2)\right\} = 0, \quad (S19)$$

where we have omitted the dependence on coupling strengths for notational convenience noting that a dependence on $X$ or $\bar{\Delta}_k$ has an implied dependence on $\bar{\Omega}_L$ or $\bar{\Omega}_k$, respectively. The function $G$ is as given in Eq. S9, and $\theta$ is as given in Eq S10 for the compensated case. Writing the coupling strengths in the form

$$\bar{\Omega}_L = 1 + q, \quad \bar{\Omega}_k = 1 + q_k,$$

it can be shown that

$$\bar{F}(X) \approx e^{-i\pi qX}, \quad (S20)$$

similar to $F(X)$, and

$$\bar{H}(X, \bar{\Delta}_1, \bar{\Delta}_2) \approx -\pi q(X - i\tfrac{\pi}{2}q) \\ - 2X\left[\left(\bar{\Delta}_1^2 + \left(\tfrac{\pi q_1}{2}\right)^2 + \left(\bar{\Delta}_1 - i\tfrac{\pi q_1}{2}\right)^2 e^{-i\phi_1}\right) \\ + \left(\bar{\Delta}_2^2 + \left(\tfrac{\pi q_2}{2}\right)^2 - \left(\bar{\Delta}_2 - i\tfrac{\pi q_2}{2}\right)^2 e^{-i\phi_2}\right)\right]. \quad (S21)$$

Hence to lowest order, we have the same result as for the compensated sequence. That is, the hyper-Ramsey pulse suppresses the shift from the uncompensated optical pulse duration. The main difference between the hyper-Ramsey sequence and compensated case comes from the additional $\bar{H}$ term which, to a good approximation, gives an additional shift.

For the dependence on the optical coupling, only the first term in Eq. S21 has a dependence on $q$ and only the real part $-\pi qX$ contributes. This gives a shift that is equal and opposite to that arising from $\bar{F}(X) \approx e^{-i\pi qX}$ and thus cancels. Moreover, the cancellation also applies to terms of order $q^2X$ that have been omitted. Subsequently, the shift for the hyper-Ramsey sequence is much more insensitive to imperfect optical coupling strengths than the compensated sequence.

For $q_1 = q_2 = 0$, the real part of the remaining terms can be written

$$-4X\left[\bar{\Delta}_1^2 \cos^2\left(\tfrac{1}{2}\phi_1\right) + \bar{\Delta}_2^2 \sin^2\left(\tfrac{1}{2}\phi_2\right)\right].$$

With $X \approx \Delta_8 \tau_L/\pi$, this provides a slight asymmetry to the magnitude of the shift as a function of detuning compared to the compensated case and this is illustrated in Fig. 6. In determining the modification to the envelope, it should be noted that maximum and minimum for the compensated case occur when $\sin \phi_k \approx \pm 1$. At these points $\cos^2(\phi_k/2) \approx \sin^2(\phi_k/2) \approx 1/2$.

As in the compensated case, non-zero values of $q_k$ will offset $\phi_k$ appearing in the above equation and replace the amplitudes $\bar{\Delta}_k^2$ by $\bar{\Delta}_k^2 + (\pi q_k/2)^2$. However the influence of this term will be dominated by the same influence $q_k$ has on $G(\bar{\Delta}_1, \bar{\Delta}_2)$ and this is the same for both the compensated and hyper-Ramsey cases.

We have not explicitly included the ac-Stark shift in the hyper-Ramsey case. To include this in Eq. S19, one simply makes the replacement

$$X \to (\delta - \Delta + \Delta_8)\tau_L/\pi \approx (-\Delta + \Delta_8)\tau_L/\pi,$$

where $\Delta$ is the effective frequency shift as given in Ref. [11], that is, the ac-Stark shift minus the compensating frequency step made during the optical pulses. Here there is an additional term $\Delta_8$. Hence the total shift is

given approximately by

$$\delta = \frac{1}{T_R}\left(\frac{4+\pi}{2}\right)\left(\frac{(\Delta - \Delta_8)\tau_L}{\pi}\right)^3 + \delta_0$$

where $\delta_0$ is the shift as previously discussed in the absence of $\Delta$. When the microwave detunings $\Delta_k$ are zero, the result is exactly that obtained in Ref. [11] noting that $\Omega_L = \pi/\tau_L$. However the factor $(4+\pi)/2$ given here is exact, where as it is specified to be approximately $\pi$ in Ref. [11]. Since the two treatments are algebraically equivalent when the microwave detunings are zero, we assume the value given in Ref. [11] to be a crude approximation.

### A. Measuring the level of suppression

It is worthwhile discussing what level of suppression can realistically be experimentally determined in light of the theoretical description. This has been explored by Physikalisch-Technische Bundesanstalt (PTB) and reported in [12], but the results are often misinterpreted as demonstrating orders of magnitude suppression. This comes from statements in the paper in reference to [12, Fig. 4b]. The statement made in the paper references the inset of the figure and notes that: if the effective light shift is kept within 0.5 Hz, the resulting clock shift would be well-below $10^{-17}$, which represents a suppression of the shift by 4 orders of magnitude. This comment is in reference to the shift $\Delta_L$ during the Ramsey pulses, that is, it compares what is effectively an uncertainty for HRS to a full shift during the optical pulses for Ramsey spectroscopy (RS). A fair comparison should compare the uncertainties for both cases, that is, to compare RS to HRS at the same value of $\Delta$, which amounts to a factor of 82. Furthermore, to justify this claim, the inset would need to include data that would demonstrate the cubic dependence with sufficient accuracy and that would require measurements with a precision of at least 10% of the vertical axis on the inset i.e. $2\times 10^{-18}$, which is 170 times smaller than the uncertainties given in their figure.

The data given is completely consistent with a straight line, which the reader can readily verify. A by-eye fit with a slope of 0.125 and no offset is certainly reasonable given that no data point is more than an uncertainty away from the line (hence the reduced-$\chi^2$ is definitely less than 1). The measurements are equally spaced by 1 Hz with an uncertainty that is approximately equal for all points at 0.25 Hz. Consequently any fitted straight line ($y = mx$) would have an uncertainty in the slope of 0.038. Hence we would estimate a slope of $m = 0.125(38)$ from the figure they present. Since the $\pi$-time for the experiment was 72 ms and the interrogation time was 144 ms, the time between the two Ramsey $\pi/2$ pulses was 216 ms. Thus conventional Ramsey spectroscopy in which the hyper-Ramsey pulse is inactive would have a slope of

$$m = \frac{1}{1 + \frac{\pi}{2}\frac{216}{72}} = 0.175$$

This is within 1.3 standard deviations of the $m = 0.125(38)$ slope we estimate from the figure. Compared to the 0.175, this would be routinely considered consistent or even excellent agreement in the scientific literature. To be absolutely clear, the data presented is completely consistent with the hyper-Ramsey pulse being inadvertently omitted (but the duration still present). This cannot seriously be suggested as a valid demonstration of a 4-order-of-magnitude suppression or an indication of any level of validity for that matter. Note that the Stark shift was 60 Hz and the 0.5 Hz requirement represents $< 1\%$ stability requirement for the laser intensity.

For our experiment 8, the ac-Stark shift during the optical pulses is 7 Hz, the interrogation time 744 ms, and the optical $\pi$-time 8 ms. In this case, the plot corresponding to [12, Fig. 4b], would have a vertical range of $\pm 0.4$ mHz or $\pm 1.1\times 10^{-18}$ fractionally for the $\pm 5$ Hz range on the horizontal axis, which is a 1000-fold smaller than for the PTB case. For the inset, the vertical range would be reduced to $\pm 8.8\times 10^{-21}$ for the $\pm 1$ Hz range shown. Hence properly validating suppression is well beyond our measurement capability. This general conclusion doesn't change for different parameter settings unless one considers extreme detunings comparable to the resolution of the optical pulses. We also note that the 0.5 Hz requirement for us is a trivially obtainable 7% stability of the laser intensity and reduces the shift to $\pm 1.1\times 10^{-21}$. However this would be compromised by experimental imperfections as discussed in the previous section, which is explicitly accounted for in our error budgets.

The best that can be done is to demonstrate a high accuracy comparison and account for deviations of the experiment from ideal parameters. That is exactly what we have done and completely consistent with what PTB has done [4, 13].

### III. MICROWAVE AC-ZEEMAN SHIFT

The shift from microwave ac-Zeeman shifts are given by

$$\delta = \frac{(\Delta_{1,7} + \Delta_{1,8})\tau_1 + (\Delta_{2,6} + \Delta_{2,7})\tau_2}{3(T + \tau_1 + \tau_2)}, \qquad \text{(S22)}$$

where $\Delta_{k,F}$ is the shift of $|F\rangle$ when $\Omega_k$ is on. Denoting the polarizations of the fields as $a_{kq}$ and the Zeeman

splitting in each hyperfine level as $\omega_F > 0$, we have

$$\Delta_{1,8} = -\frac{\Omega_{8,0,7,1}^2}{4\omega_7} + \frac{\Omega_{8,0,7,-1}^2}{4\omega_7},$$

$$= -\frac{7}{16}\frac{\Omega_{8,0,7,0}^2}{4\omega_7}\left[\left(\frac{a_{1+}}{a_{10}}\right)^2 - \left(\frac{a_{1-}}{a_{10}}\right)^2\right],$$

$$\Delta_{1,7} = \frac{\Omega_{8,-1,7,0}^2}{4\omega_8} - \frac{\Omega_{8,1,7,0}^2}{4\omega_8},$$

$$= \frac{9}{16}\frac{\Omega_{8,0,7,0}^2}{4\omega_8}\left[\left(\frac{a_{1+}}{a_{10}}\right)^2 - \left(\frac{a_{1-}}{a_{10}}\right)^2\right],$$

$$\Delta_{2,7} = \frac{\Omega_{7,0,6,1}^2}{4\omega_6} - \frac{\Omega_{7,0,6,-1}^2}{4\omega_6},$$

$$= \frac{3}{7}\frac{\Omega_{7,0,6,0}^2}{4\omega_6}\left[\left(\frac{a_{2+}}{a_{20}}\right)^2 - \left(\frac{a_{2-}}{a_{20}}\right)^2\right],$$

$$\Delta_{2,6} = \frac{\Omega_{7,-1,6,0}^2}{4\omega_7} - \frac{\Omega_{7,1,6,0}^2}{4\omega_7},$$

$$= \frac{4}{7}\frac{\Omega_{7,0,6,0}^2}{4\omega_7}\left[\left(\frac{a_{2+}}{a_{20}}\right)^2 - \left(\frac{a_{2-}}{a_{20}}\right)^2\right].$$

We measure the polarization ratios at a fixed (high) power and the ratio of $\pi$ times is determines the ratio of the polarization components. Hence

$$\frac{a_{1\pm}}{a_{10}} = \frac{4}{3}\frac{\tau_{10}}{\tau_{1\pm}}, \quad \frac{a_{2\pm}}{a_{20}} = \sqrt{\frac{7}{3}}\frac{\tau_{20}}{\tau_{2\pm}},$$

where $\tau_{kq}$ is the $\pi$-time for the polarization $q$ of field $k$ and we have omitted signs from Clebsch-Gordan factors. Thus we have

$$\Delta_{1,8} + \Delta_{1,7} = \left(-\frac{7}{9}\frac{\Omega_{8,0,7,0}^2}{4\omega_7} + \frac{\Omega_{8,0,7,0}^2}{4\omega_8}\right)Y_1,$$

$$\Delta_{2,7} + \Delta_{2,6} = \left(\frac{\Omega_{7,0,6,0}^2}{4\omega_6} + \frac{4}{3}\frac{\Omega_{7,0,6,0}^2}{4\omega_7}\right)Y_2.$$

where

$$Y_k = \left(\frac{\tau_{k0}}{\tau_{k+}}\right)^2 - \left(\frac{\tau_{k0}}{\tau_{k-}}\right)^2.$$

The total error is dominated by $\Delta_{1,8}$ and $\Delta_{2,6}$ due to the smaller size of the Zeeman splitting $\omega_7$. The factors in parentheses are given by the microwave $\pi$ times and the magnetic field used in the experiment. For a $\pi$-time of $\tau_r = 1$ ms and a magnetic field of $B_r = 0.1$ mT, we have

$$a_{1r} = \frac{1}{2\pi f_c}\left(-\frac{7}{9}\frac{\Omega_{8,0,7,0}^2}{4\omega_7} + \frac{\Omega_{8,0,7,0}^2}{4\omega_8}\right)$$
$$\approx -9.62 \times 10^{-15}, \quad (S23)$$

$$a_{2r} = \frac{1}{2\pi f_c}\left(\frac{\Omega_{7,0,6,0}^2}{4\omega_6} + \frac{4}{3}\frac{\Omega_{7,0,6,0}^2}{4\omega_7}\right)$$
$$\approx 2.18 \times 10^{-14}. \quad (S24)$$

These values can then be scaled for any experiment using the specific values of $\tau$ and $T_R$ used giving a total fractional shift of

$$\frac{\tau_r^2 B_r}{\tau T_R B}\left(a_{1r}Y_1 + a_{2r}Y_2\right).$$

Uncertainties are dominated by the determination of $Y_k$, as the associated $\pi$-times are measured less accurately. Taking the same fractional uncertainty $\beta$ for the $\pi$-times, we have

$$\delta Y_k = \beta\sqrt{\frac{8\tau_0^4(\tau_+^4 - \tau_+^2\tau_-^2 + \tau_-^4)}{\tau_-^4\tau_+^4}},$$

where we have omitted the subscript $k$ for convenience. If we write $\tau_{\min} = \min(\tau_+, \tau_-)$ and $\tau_{\max} = \max(\tau_+, \tau_-) = \gamma\tau_{\min}$, then the above expression can be written

$$\beta\left(\frac{\tau_0}{\tau_{\min}}\right)^2\sqrt{8\frac{\gamma^4 - \gamma^2 + 1}{\gamma^4}}.$$

The last term lies between $\sqrt{6}$ and $\sqrt{8}$ so is within 7% of $2.64 \approx \sqrt{7}$. Although the shift is suppressed when the circular polarizations are balanced, the uncertainty is primarily determined by their suppression.

Contributions from different $k$ would add in quadrature giving a fractional uncertainty of

$$\frac{\tau_r^2 B_r}{\tau T_R B}\sqrt{a_{1r}^2\delta Y_1^2 + a_{2r}^2\delta Y_2^2}.$$

We are then able to write the fractional uncertainty for LC-$n$ as

$$\delta_n = \alpha_n\frac{\tau_r^2 B_r}{\tau T_R B}. \quad (S25)$$

From the measured $\pi$-times for the characterizations of the polarizations, $\alpha_n$ differs from the mean values by only $\pm 10\%$ and we have

$$\alpha_1 \approx 7.1 \times 10^{-16}, \quad \text{and} \quad \alpha_2 \approx 3.0 \times 10^{-15}.$$

The larger value for LC-2 (by about a factor of 4) is a consequence of the fact that the microwave polarizations for the $|7\rangle \leftrightarrow |6\rangle$ were approximately equal i.e. the circular components had the worst suppression, which happened to be for the largest contribution to the error. The worst case is for experiment 5 which would have an uncertainty contribution of $\sim 7 \times 10^{-19}$ for LC-2 (at low field). For experiments 7 and 8 the uncertainty is $1 \sim 2 \times 10^{-19}$.

## IV. THE BBR SHIFT

Given the very low BBR shift for the $^1S_0 \leftrightarrow {}^3D_1$ transition, it is possible to make a very crude estimate to bound the shift. Our traps are four rod traps with

axial end caps and all electrodes are 0.48mm diameter rods of copper-beryllium. At each end of the trap the 5 electrodes (4 rods + end cap) are fixed in place by a $4 \times 4 \times 3$ mm cube of alumina press fit into a larger copper block ($\sim 9 \times 9 \times 9$ mm). Each copper block, top and bottom is attached to a 2mm thick copper plate which would act as an effective heat spreader. One of the plates is directly bolted to a custom outer aluminium cylinder that is bolted directly to the vacuum chamber wall. The RF rods are directly clamped to a 3 mm diameter, 5cm long, rf copper feedthrough pin.

Taking the thermal conduction in the alumina spacer only over the 90 degree angle between vertical and horizontal, the thermal resistance is $\sim 5$ K/W. There are two rods and two alumina spacers, so taking the resistance in parallel we would have $\sim 1$ K/W thermal resistance. Taking the worst-case scenario of all heat conducted along the full length of one copper plate gives an added thermal resistance of 3 K/W and we estimate a further 1.5 K/W for the outer cylinder giving $\lesssim 5.5$ K/W thermal resistance from electrode to chamber wall. This neglects any radiation losses from any surface, any conduction losses from the direct RF pin connection, and the heat spreading effect of the second copper plate. The input power for our experiment is $\sim 2$ W. Assuming all this is dissipated in the RF electrodes gives at most a 11 K temperature rise above the lab environment, which was no less than 25 °C or 298 K during these experiments. We are aware of systems in which there has been observed much larger temperature rises such as [14], but these systems have considerably larger input powers than we use.

All things considered, we think a 35(10) °C bound on the temperature is a conservative estimate inclusive of the worst case scenario. However, we refrain from making accuracy claims on the basis of this argument. We also stress that the comparison reported concerns the temperature difference, which is highly insensitive to an overall absolute temperature. As noted in the main text, the temperature difference would need to be significantly more than 20°C to have any bearing on the result.

## V.  UNCERTAINTY BUDGETS FOR EXPERIMENTS 1-6

Table I gives the measured frequency differences between Lu-1 and Lu-2 for experiments 1 to 6, with statistical and systematic uncertainties given separately. Tables II-VII give the uncertainty budgets for experiments 1 to 6.

TABLE I. Measured clock frequency differences in mHz for experiments 1-7. 'Stat.' and 'Sys.' are the statistical and systematic uncertainties respectively in mHz. $B_k$ are the average magnetic fields in Lu-$k$.

| Expt. | Diff. | Stat. | Sys. | $B_1$(mT) | $B_2$(mT) |
|---|---|---|---|---|---|
| 1 | −17.5177 | 7.4 | 12.1 | 1.894286 | 0.101049 |
| 2 | −17.5130 | 2.8 | 6.7 | 1.894212 | 0.101040 |
| 3 | −17.5012 | 3.2 | 3.8 | 1.894030 | 0.101005 |
| 4 | −17.5065 | 3.9 | 6.8 | 1.894055 | 0.100967 |
| 5 | −17.4992 | 5.3 | 4.0 | 1.894009 | 0.100926 |
| 6 | −7.8431 | 2.1 | 3.8 | 1.270804 | 0.100552 |

TABLE II. Uncertainty budget for Experiment 1 in which $\tau_L = 4.4\,\text{ms}, \tau_1 = \tau_2 = 12\,\text{ms}, T = 100\,\text{ms}, B_1 \approx 1.9\,\text{mT}, B_2 \approx 0.1\,\text{mT}$. All values are relative to $10^{-18}$ of the HA frequency.

| Effect | Lu-1 Shift | Lu-1 Unc. | Lu-2 Shift | Lu-2 Unc. | Difference Shift | Difference Unc. |
|---|---|---|---|---|---|---|
| Excess micromotion | -0.41 | 0.37 | -0.44 | 0.34 | 0.03 | 0.50 |
| Second-order Doppler (thermal) | -1.32 | 0.31 | -0.10 | 0.06 | -1.22 | 0.32 |
| ac-Zeeman (rf) | 0.54 | 0.01 | 0.15 | 0.01 | 0.39 | 0.01 |
| ac-Zeeman (microwave) | -0.03 | 0.01 | -0.22 | 0.60 | 0.19 | 0.60 |
| Gravity shift | - | - | - | - | -1.31 | 0.15 |
| ac-Stark shift | -485.37 | 24.27 | -474.83 | 23.74 | -10.54 | 33.95 |
| HARS timing | 9.27 | 2.84 | 0.90 | 0.79 | 8.37 | 2.95 |
| Microwave coupling | 0.00 | 0.42 | 0.00 | 0.42 | 0.00 | 0.60 |
| Optical coupling | 0.00 | 0.30 | 0.00 | 0.30 | 0.00 | 0.43 |
| Residual quadruple shift | 0.22 | 0.02 | 0 | 0.32 | 0.22 | 0.32 |
| **Total** | -477.09 | 24.44 | -474.54 | 23.77 | -3.87 | 34.10 |

TABLE III. Uncertainty budget for Experiment 2 in which $\tau_L = 4.4\,\text{ms}, \tau_1 = \tau_2 = 12\,\text{ms}, T = 200\,\text{ms}, B_1 \approx 1.9\,\text{mT}, B_2 \approx 0.1\,\text{mT}$. All values are relative to $10^{-18}$ of the HA frequency.

| Effect | Lu-1 Shift | Lu-1 Unc. | Lu-2 Shift | Lu-2 Unc. | Difference Shift | Difference Unc. |
|---|---|---|---|---|---|---|
| Excess micromotion | -0.41 | 0.37 | -0.44 | 0.34 | 0.03 | 0.50 |
| Second-order Doppler (thermal) | -1.77 | 0.42 | -0.12 | 0.06 | -1.64 | 0.42 |
| ac-Zeeman (rf) | 0.54 | 0.01 | 0.15 | 0.01 | 0.39 | 0.01 |
| ac-Zeeman (microwave) | -0.02 | 0.00 | -0.12 | 0.33 | 0.10 | 0.33 |
| Gravity shift | - | - | - | - | -1.31 | 0.15 |
| ac-Stark shift | -268.68 | 13.43 | -262.85 | 13.14 | -5.83 | 18.79 |
| HARS timing | 12.28 | 1.58 | 0.54 | 0.44 | 11.73 | 1.64 |
| Microwave coupling | 0.00 | 0.23 | 0.00 | 0.23 | 0.00 | 0.33 |
| Optical coupling | 0.00 | 0.17 | 0.00 | 0.17 | 0.00 | 0.24 |
| Residual quadruple shift | 0.22 | 0.02 | 0 | 0.32 | 0.22 | 0.32 |
| **Total** | -257.84 | 13.54 | -262.84 | 13.17 | 3.69 | 18.89 |

TABLE IV. Uncertainty budget for Experiment 3 in which $\tau_L = 8\,\text{ms}, \tau_1 = \tau_2 = 12\,\text{ms}, T = 200\,\text{ms}, B_1 \approx 1.9\,\text{mT}, B_2 \approx 0.1\,\text{mT}$. All values are relative to $10^{-18}$ of the HA frequency.

| Effect | Lu-1 Shift | Lu-1 Unc. | Lu-2 Shift | Lu-2 Unc. | Difference Shift | Difference Unc. |
|---|---|---|---|---|---|---|
| Excess micromotion | -0.41 | 0.37 | -0.44 | 0.34 | 0.03 | 0.50 |
| Second-order Doppler (thermal) | -1.77 | 0.42 | -0.12 | 0.06 | -1.64 | 0.42 |
| ac-Zeeman (rf) | 0.54 | 0.01 | 0.15 | 0.01 | 0.39 | 0.01 |
| ac-Zeeman (microwave) | -0.02 | 0.00 | -0.12 | 0.33 | 0.10 | 0.33 |
| Gravity shift | - | - | - | - | -1.31 | 0.15 |
| ac-Stark shift | -146.48 | 7.32 | -144.23 | 7.21 | -2.25 | 10.28 |
| HARS timing | 7.13 | 2.86 | 1.31 | 0.80 | 5.82 | 2.97 |
| Microwave coupling | 0.00 | 0.23 | 0.00 | 0.23 | 0.00 | 0.33 |
| Optical coupling | 0.00 | 0.31 | 0.00 | 0.31 | 0.00 | 0.43 |
| Residual quadruple shift | 0.22 | 0.02 | 0 | 0.32 | 0.22 | 0.32 |
| **Total** | -140.78 | 7.89 | -143.45 | 7.29 | 1.37 | 10.74 |

TABLE V. Uncertainty budget for Experiment 4 in which $\tau_L = 8\,\text{ms}, \tau_1 = \tau_2 = 12\,\text{ms}, T = 100\,\text{ms}, B_1 \approx 1.9\,\text{mT}, B_2 \approx 0.1\,\text{mT}$. All values are relative to $10^{-18}$ of the HA frequency.

| Effect | Lu-1 Shift | Lu-1 Unc. | Lu-2 Shift | Lu-2 Unc. | Difference Shift | Difference Unc. |
|---|---|---|---|---|---|---|
| Excess micromotion | -0.41 | 0.37 | -0.44 | 0.34 | 0.03 | 0.50 |
| Second-order Doppler (thermal) | -1.32 | 0.31 | -0.10 | 0.06 | -1.22 | 0.32 |
| ac-Zeeman (rf) | 0.54 | 0.01 | 0.15 | 0.01 | 0.39 | 0.01 |
| ac-Zeeman (microwave) | -0.03 | 0.01 | -0.22 | 0.60 | 0.19 | 0.60 |
| Gravity shift | - | - | - | - | -1.31 | 0.15 |
| ac-Stark shift | -264.61 | 13.23 | -260.54 | 13.03 | -4.06 | 18.57 |
| HARS timing | 34.70 | 5.14 | 2.99 | 1.44 | 31.72 | 5.34 |
| Microwave coupling | 0.00 | 0.42 | 0.00 | 0.42 | 0.00 | 0.60 |
| Optical coupling | 0.00 | 0.55 | 0.00 | 0.55 | 0.00 | 0.78 |
| Residual quadruple shift | 0.22 | 0.02 | 0 | 0.32 | 0.22 | 0.32 |
| **Total** | -230.90 | 14.22 | -258.16 | 13.15 | 25.95 | 19.37 |

TABLE VI. Uncertainty budget for Experiment 5 in which $\tau_L = 8\,\text{ms}, \tau_1 = \tau_2 = 6\,\text{ms}, T = 200\,\text{ms}, B_1 \approx 1.9\,\text{mT}, B_2 \approx 0.1\,\text{mT}$. All values are relative to $10^{-18}$ of the HA frequency.

| Effect | Lu-1 Shift | Lu-1 Unc. | Lu-2 Shift | Lu-2 Unc. | Difference Shift | Difference Unc. |
|---|---|---|---|---|---|---|
| Excess micromotion | -0.41 | 0.37 | -0.44 | 0.34 | 0.03 | 0.50 |
| Second-order Doppler (thermal) | -1.71 | 0.41 | -0.12 | 0.06 | -1.59 | 0.41 |
| ac-Zeeman (rf) | 0.54 | 0.01 | 0.15 | 0.01 | 0.39 | 0.01 |
| ac-Zeeman (microwave) | -0.03 | 0.01 | -0.25 | 0.70 | 0.22 | 0.70 |
| Gravity shift | - | - | - | - | -1.31 | 0.15 |
| ac-Stark shift | -154.77 | 7.74 | -152.39 | 7.62 | -2.38 | 10.86 |
| HARS timing | 19.17 | 3.02 | 2.16 | 0.84 | 17.01 | 3.14 |
| Microwave coupling | 0.00 | 0.25 | 0.00 | 0.25 | 0.00 | 0.35 |
| Optical coupling | 0.00 | 0.32 | 0.00 | 0.32 | 0.00 | 0.46 |
| Residual quadruple shift | 0.22 | 0.02 | 0 | 0.32 | 0.22 | 0.32 |
| **Total** | -137.00 | 8.34 | -150.90 | 7.72 | 12.59 | 11.37 |

TABLE VII. Uncertainty budget for Experiment 6 in which $\tau_L = 8\,\text{ms}, \tau_1 = \tau_2 = 12\,\text{ms}, T = 200\,\text{ms}, B_1 \approx 1.2\,\text{mT}, B_2 \approx 0.1\,\text{mT}$. All values are relative to $10^{-18}$ of the HA frequency.

| Effect | Lu-1 Shift | Lu-1 Unc. | Lu-2 Shift | Lu-2 Unc. | Difference Shift | Difference Unc. |
|---|---|---|---|---|---|---|
| Excess micromotion | -0.41 | 0.37 | -0.44 | 0.34 | 0.03 | 0.50 |
| Second-order Doppler (thermal) | -1.77 | 0.42 | -0.12 | 0.06 | -1.64 | 0.42 |
| ac-Zeeman (rf) | 0.54 | 0.01 | 0.15 | 0.01 | 0.39 | 0.01 |
| ac-Zeeman (microwave) | -0.02 | 0.01 | -0.12 | 0.33 | 0.10 | 0.33 |
| Gravity shift | - | - | - | - | -1.31 | 0.15 |
| ac-Stark shift | -153.45 | 7.67 | -140.80 | 7.04 | -12.64 | 10.41 |
| HARS timing | -6.12 | 1.84 | 5.53 | 0.80 | -11.64 | 2.00 |
| Microwave coupling | 0.00 | 0.23 | 0.00 | 0.23 | 0.00 | 0.33 |
| Optical coupling | 0.00 | 0.31 | 0.00 | 0.31 | 0.00 | 0.43 |
| Residual quadruple shift | 0.22 | 0.02 | 0 | 0.32 | 0.22 | 0.32 |
| **Total** | -161.00 | 7.92 | -135.80 | 7.12 | -26.50 | 10.65 |


\end{document}